\documentclass[11pt]{article}

\usepackage{amsmath}
\usepackage{amssymb}
\usepackage{graphicx}
\usepackage{dcolumn}
\usepackage{bm}
\usepackage{lscape}
\usepackage{tabularx}
\usepackage[titles]{tocloft}

\newcommand{\append}[1]{\protect\refstepcounter{section}
                \section*{Appendix \thesection \, #1}
                \addcontentsline{toc}{section}{Appendix \thesection: #1}}

\begin{document}

\begin{flushright}
CERN-PH-TH/2011-135\\
LYCEN 2011-06 
\end{flushright}

\begin{center}
\Large\bf\boldmath
\vspace*{2.0cm}AlterBBN: A program for calculating the BBN abundances\\
of the elements in alternative cosmologies 
\unboldmath
\end{center}

\vspace{0.8cm}
\begin{center}
Alexandre Arbey\footnote{\tt alexandre.arbey@ens-lyon.fr\\
\hspace*{0.6cm}URL: \tt http://superiso.in2p3.fr/relic/alterbbn}\\[0.4cm]
{\sl Universit\'e de Lyon, France; Universit\'e Lyon 1, CNRS/IN2P3,\\ UMR5822 IPNL, F-69622~Villeurbanne Cedex, France;\\ Centre de Recherche Astrophysique de Lyon, Observatoire de Lyon, Saint-Genis Laval cedex, F-69561; CNRS, UMR 5574; Ecole Normale Sup\'erieure de Lyon, Lyon, France.}\\
\vspace{0.5cm}
{\sl CERN Theory Division, Physics Department, CH-1211 Geneva 23, Switzerland}\\
\end{center}
\vspace{0.9cm}
\begin{abstract}
\noindent We describe \verb?AlterBBN?, a public C program for evaluating the abundances of the elements generated by Big-Bang nucleosynthesis (BBN). This program enables the user to compute the abundances of the elements in the standard model of cosmology, and additionally provides possibilities to alter the assumptions of the cosmological model in order to study their consequences on the abundances of the elements. In particular the baryon-to-photon ratio and the effective number of neutrinos, as well as the expansion rate and the entropy content of the Universe during BBN can be modified in \verb?AlterBBN?. Such features allow the user to test the cosmological models by confronting them to BBN constraints. A presentation of the physics of BBN and the features of \verb?AlterBBN? is provided here under the form of a manual.
\end{abstract}
\newpage
\tableofcontents
%
\newpage
\section{Introduction} 
Big Bang Nucleosynthesis (BBN) is an important part of the Cosmological Standard Model, as it gave birth to the first nuclei. Based on BBN models it is possible to calculate the abundances of the light elements in the Universe, which can then be compared to the measured abundances. This allows to probe the Universe properties at the time of nucleosynthesis, which is the most ancient period observationally accessible. In standard cosmology, only one free parameter, the baryon-to-photon ratio at BBN time is needed to determine the abundances of the elements, and this parameter can be deduced from the observation of the Cosmic Microwave Background \cite{wmap}. To calculate the abundances of the elements in the standard model of cosmology, several ``semi-public'' or private codes, such as NUC123 \cite{kawano} or PArthENoPE \cite{parthenope}, already exist.

We provide here a public code, \verb?AlterBBN?, which enables the calculation of the abundances of the elements in the standard cosmological model, as well as in alternative cosmological models such as quintessence or reheating scenarios. Indeed, many different phenomena in the Early Universe could have modified the Universe properties at the time of BBN, and in such cases the calculations could lead to abundances of the elements different from the predictions in the standard cosmological model. One of the main features of \verb?AlterBBN? is that it provides calculations in such cosmological models, which can enable the user to test alternative scenarios and to constrain their parameters.

The code \verb?AlterBBN? has been inspired by NUC123, and we use similar calculation techniques throughout the program. In the following, we describe first the physics of Big-Bang nucleosynthesis, and present the set of equations in the standard model of cosmology. We then discuss alternative cosmological models, with a modified expansion rate or entropy production. We describe also the content of the \verb?AlterBBN? package, as well as the list of its main routines. We will then explain how to compile and use \verb?AlterBBN?, and the inputs and outputs of the program will be introduced. Finally, some examples of results obtained with \verb?AlterBBN? will be given. In the Appendices, we give the list of the nuclear reactions used in \verb?AlterBBN?, as well as experimental limits which can be considered to constrain the calculated abundances of the elements.

\section{BBN physics}
In the following section, we use the natural unit system $c = \hbar = k =1$.
\subsection{BBN in the Cosmological Standard Model}
\label{costamo}
All the nuclear reactions implemented in \verb?AlterBBN? have evolution equations of the form:
\begin{equation}
n_i\,^{A_i}\!Z_i \,+\, n_j\,^{A_j}\!Z_j \,\longleftrightarrow\, n_k\,^{A_k}\!Z_k \,+\, n_l\,^{A_l}\!Z_l\;.
\end{equation}
Therefore, the abundance change for nuclide $i$ is given by \cite{kawano}:
\begin{equation}
\frac{dY_i}{dt}=\sum_{j,k,l} n_i \left(-\frac{Y_i^{n_i} Y_j^{n_j}}{n_i!\, n_j!} \Gamma_{ij}^k + \frac{Y_l^{n_l} Y_k^{n_k}}{n_l!\, n_k!} \Gamma_{lk}^j \right) \;,
\end{equation}
where the nuclide abundance is $Y_i = X_i / A_i$, $X_i$ being the mass fraction in nuclide $i$ and $A_i$ its atomic number. $n_i$ is the number of nuclides $i$ involved in the reaction, $\Gamma_{ij}^k$ is the forward reaction rate and $\Gamma_{lk}^j$ the reverse rate.\\
The expansion rate of the Universe is given by the Friedmann equation:
\begin{equation}
H^2=\left(\frac{\dot{a}}{a}\right)=\frac{8\pi G}{3} \rho_{tot} \;,
\end{equation}
and the equation of energy conservation can be written as:
\begin{equation}
\frac{d}{dt}(\rho_{tot} a^3) + P_{tot} \frac{d}{dt}(a^3) - a^3 \left.\frac{d\rho_{tot}}{dt}\right|_{T=cst} = 0\;,
\end{equation}
where
\begin{equation}
\rho_{tot} = \rho_\gamma + \rho_\nu  + \rho_b + (\rho_{e^-}+\rho_{e^+}) \;,
\end{equation}
and
\begin{equation}
P_{tot} = P_\gamma + (P_{e^-}+P_{e^+}) + P_b \;.
\end{equation}
We have
\begin{equation}
P_\gamma = \frac13 \rho_\gamma \;, \qquad \rho_\gamma=\frac{\pi}{15} T^4\;,
\end{equation}
and in absence of neutrino degeneracy:
\begin{equation}
\rho_\nu= N_\nu \frac78 \,\frac{\pi}{15} T_\nu^4 = N_\nu \frac{7\pi}{120} \left(\frac{4}{11}\right)^{4/3} T^4\;,
\end{equation}
where $N_\nu=3$ is the number of neutrino families. The sum of the electron and positron densities is
\begin{equation}
\rho_{e^-} + \rho_{e^+} = \frac{2}{\pi^2} m_e^4 \sum_{n=1}^\infty \frac{(-1)^{n+1}}{nz}\cosh(n \phi_e) M(nz)\;,
\end{equation}
and the sum of their pressures is
\begin{equation}
P_{e^-} + P_{e^+} = \frac{2}{\pi^2} m_e^4 \sum_{n=1}^\infty \frac{(-1)^{n+1}}{nz}\cosh(n \phi_e) L(nz)\;,
\end{equation}
where $z=m_e/T$,
\begin{equation}
L(z)=\frac{K_2(z)}{z} \;, \qquad M(z)=\frac1z \left[\frac34 K_3(z) + \frac14 K_1(z) \right] \;, \qquad N(z)=\frac1z \left[\frac12 K_4(z) + \frac12 K_2(z) \right] \;,
\end{equation}
the $K_i$ being the modified Bessel functions. The charge conservation gives
\begin{equation}
n_{e^-} - n_{e^+} = \mathcal{N}_A h T^3 S \;,
\end{equation}
where $\mathcal{N}_A$ is the Avogadro number,
\begin{equation}
S = \sum_i Z_i Y_i \;,
\end{equation}
and the difference between the electron and positron densities is
\begin{equation}
n_{e^-} - n_{e^+} = \frac{2}{\pi^2} m_e^3 \sum_{n=1}^\infty (-1)^{n+1}\sinh(n \phi_e) L(nz) \;,
\end{equation}
from which $\phi_e$ can be determined using:
\begin{equation}
 \frac{d\phi_e}{dt} = \frac{\partial \phi_e}{\partial T} \frac{dT}{dt} + \frac{\partial \phi_e}{\partial a} \frac{da}{dt} + \frac{\partial \phi_e}{\partial S} \frac{dS}{dt} \;.
\end{equation}
The baryon energy density reads
\begin{equation}
\rho_b = h T^3 \left[1 + \sum_i \left( \frac{\Delta M_i}{M_u} + \zeta T \right) Y_i \right] \;,
\end{equation}
where $\Delta M_i$ is the mass excess of nuclide $i$, and the pressure is
\begin{equation}
P_b = h T^3 \left( \frac23 \zeta T \sum_i Y_i \right) \;,
\end{equation}
with $\zeta=1.388 \times 10^{-4}$. $h \sim \rho_b / T^3 \sim 1 / a^3 / T^3$ can be determined by the equation:
\begin{equation}
\frac{dh}{dt}=-3 h \left( \frac1a \frac{da}{dt} + \frac1T \frac{dT}{dt}\right)\;.
\end{equation}
To solve the system of differential equations which gives the abundances of the elements, we assume the following initial conditions:
\begin{equation}
h(T_i) = M_u \frac{n_\gamma(T_i)}{T_i^3} \eta, 
\end{equation}
where $M_u$ is the unit atomic mass, $n_\gamma$ is the number density of photons and $\eta$ the baryon-to-photon ratio.
Also,
\begin{equation}
\phi_e(T_i) \approx \frac{\pi^2}{2} \mathcal{N}_A \frac{h(T_i) Y_p}{z_i^3} \frac{1}{\sum_{n=1}^\infty (-1)^{n+1} n L(n z_i)} \;,
\end{equation}
and the initial abundances of protons and neutrons are
\begin{equation}
 Y_p(T_i) = \frac{1}{1+e^{-q/T_i}} \;, \qquad Y_n(T_i) = \frac{1}{1+e^{q/T_i}}\;,
\end{equation}
where $q=m_n-m_p$. The other quantities needed at this point are the initial baryon energy density
\begin{equation}
\rho_b(T_i) \approx h(T_i) T_i^3 \;,
\end{equation}
and the time
\begin{equation}
t_i = (12 \pi G \sigma)^{1/2} T_i^{-2} \;,
\end{equation}
where $\sigma$ is the Stefan-Boltzmann constant.\\
\\
To solve this system of equations, a linearization of the nuclear abundance differential equations is performed, followed by a Runge-Kutta integration.
%
\subsection{Modified Cosmological Scenarios}%
\label{mocomo}
The Early Universe is a relatively unknown period, and therefore the properties of the Universe at that period are not well-known either. We consider in the following several alternative descriptions of the Early Universe properties.
%

\subsubsection{Modified expansion rate}
\label{modexprate}%
Many phenomena at the time of BBN could have effects on the expansion rate of the Universe. In such cases, the Friedmann equation is affected, and we parametrize the modification of the expansion by adding an effective dark energy density $\rho_D$ to the total energy of the Universe. The Friedmann equation then reads:
\begin{equation}
H^2=\left(\frac{\dot{a}}{a}\right)=\frac{8\pi G}{3} (\rho_{tot}+\rho_D) \;.
\end{equation}
We adopt the parametrizations described in \cite{arbey,arbey2} for $\rho_D$:
\begin{equation}
 \rho_D =  \kappa_\rho \rho_{tot}(T_0) \left(\frac{T}{T_0}\right)^{n_\rho} \label{rhoD} \;,
\end{equation}
where $T_0 = 1$ MeV. $\kappa_\rho$ is therefore the ratio of effective dark energy density over total energy density, and $n_\rho$ is a parameter describing the behavior of this density. We refer the reader to \cite{arbey,arbey2} for detailed descriptions and discussions of this parametrization.

\subsubsection{Modified entropy content}
\label{modentropy}%
The entropy content can also receive various contributions in the BBN period. We parametrize the contributions by considering an effective dark entropy density $s_D$. The energy conservation equation reads in this case:
\begin{equation}
\frac{d}{dt}(\rho_{tot} a^3) + P_{tot} \frac{d}{dt}(a^3) - a^3 \left.\frac{d\rho_{tot}}{dt}\right|_{T=cst} - T \frac{d}{dt}(s_{D} a^3) = 0\;.
\end{equation}
We adopt two different parametrization for $s_D$. The first one has been introduced in \cite{arbey2}:
\begin{equation}
s_D =  \kappa_s s_{rad}(T_0) \left(\frac{T}{T_0}\right)^{n_s} \;,\label{sD}
\end{equation}
where $T_0 =1$ MeV and
\begin{equation}
s_{rad}(T)=h_{\mbox{eff}}(T) \frac{2 \pi^2}{45} T^3 \;,
\end{equation}
$h_{\mbox{eff}}$ being the effective number of entropic degrees of freedom of radiation, $\kappa_s$ the ratio of effective dark entropy density over radiation entropy density, and $n_s$ a parameter describing the behavior of this density. We refer the reader to \cite{arbey2} for detailed descriptions and discussions of this parametrization.\\
\\
The second parametrization is inspired by reheating scenarios in which the entropy production $\Sigma_D$ evolves like \cite{arbey3}
\begin{equation}
\Sigma_D(T) = \kappa_\Sigma \Sigma_{rad}(T_0) \left(\frac{T}{T_0}\right)^{n_\Sigma}
\end{equation}
where $T_0 = 1$ MeV. $\kappa_\Sigma$ is the ratio of effective dark entropy production over radiation entropy production, and $n_\Sigma$ is a parameter describing the behavior of this entropy production ($n_\Sigma \sim -1$ in most reheating scenarios). The radiation entropy production reads:
\begin{equation}
\Sigma_{rad}(T_0) = \left(\frac{4\pi^3G}{5} g_{\mbox{eff}}(T_0)\right)^{1/2} T_0^2 s_{rad}(T_0)\;,
\end{equation}
$g_{\mbox{eff}}$ being the effective number of degrees of freedom of radiation, and where the entropy production stops at $T=T_r$. The function $\Sigma_D(T)$ is related to the entropy density by:
\begin{equation}
 s_D(T) = 3 \sqrt{\frac{5}{4 \pi^3 G}} h_{\mbox{eff}} T^3 \int_0^T dT' \frac{g_*^{1/2}\Sigma_D(T')}{\sqrt{1+\dfrac{\rho_D}
 {\rho_{rad}}}h_{\mbox{eff}}^2(T') T'^6}\;,
\label{sigmas}
\end{equation}
where
\begin{equation}
g_*^{1/2}=\frac{h_{\mbox{eff}}}{\sqrt{g_{\mbox{eff}}}}\left(1+\frac{T}{3 h_{\mbox{eff}}}\frac{dh_{\mbox{eff}}}{dT}\right) \;.
\end{equation}
This parametrization is further described in \cite{arbey3}.
%
\section{Content of the AlterBBN package}%
\label{content}
\verb?AlterBBN? is based on the library \verb?libbbn.a?, which contains all the procedure necessary to the abundance computation, and is provided with five different programs.

\subsection{Parameter structure}
The main structure of the package \verb?AlterBBN? is defined in \verb?src/include.h? and is provided below:
\begin{verbatim}
typedef struct relicparam
{
	int entropy_model;
	double dd0,ndd,Tdend;
	double sd0,nsd,Tsend;
	double Sigmad0,nSigmad,TSigmaend;
	double nt0,nnt,Tnend;
	double mgravitino;
	double table_eff[276][3];
	double eta0;
	double nbnu;
	double life_neutron;
}
relicparam;
\end{verbatim}
This structure contains all the parameters which are necessary to the computation of the abundances of the elements in the standard cosmological model as well as in alternative models.
\subsection{Main routines}
\label{mainroutines}
The main routines of the \verb?libbbn.a? library, which are needed for the abundance calculation, are detailed in the following:
\begin{itemize}
\item \verb?void Init_cosmomodel(struct relicparam* paramrelic)?\\
\\
This function initializes the \verb?paramrelic? structure, setting all the parameters to 0, apart from the number of neutrinos which is set to 3, and the baryon-to-photon ratio $\eta=6.19 \times 10^{-10}$ \cite{wmap}.\\
\item \verb?void Init_cosmomodel_param(double eta, double nbnu, struct relicparam* paramrelic)?\\
\\
This routine specifies values of the \verb?paramrelic? structure, enabling the user to modify the baryon-to-photon ratio {\tt eta} and the number of neutrinos {\tt nbnu}.\\
\item \verb?void Init_dark_density(double dd0, double ndd, double T_end, struct relicparam*?\\
\verb?paramrelic)?\\
\\
This routine changes values in the \verb?paramrelic? structure, initializing the effective dark density parameters described in Appendix~\ref{modexprate}, with $\kappa_\rho$={\tt dd0}, $n_\rho$={\tt ndd}, and where \verb?T_end? is a temperature cut at which the dark density is set to 0.\\

\item \verb?void Init_dark_entropy(double sd0, double nsd, double T_end, struct relicparam*?\\
\verb?paramrelic)?\\
\\
This routine changes values in the \verb?paramrelic? structure, initializing the effective dark entropy density parameters described in Appendix~\ref{modentropy}, with $\kappa_s$={\tt sd0}, $n_s$={\tt nsd}, and where \verb?T_end? is a temperature cut at which the dark entropy density is set to 0.\\

\item \verb?void Init_dark_entropySigmaD(double Sigmad0, double nSigmad, double T_end,?\\
\verb?struct relicparam* paramrelic)?\\
\\
This routine specifies values of the \verb?paramrelic? structure, initializing the effective dark entropy production parameters described in Appendix~\ref{modentropy}, with $\kappa_\Sigma$={\tt Sigmad0}, $n_\Sigma$={\tt nSigmad}, and where \verb?T_end? is a temperature cut at which the dark entropy production is set to 0.\\

\item \verb?void rate_weak(double f[])?\\
\\
This procedure computes the decay rates of the weak interaction reactions, and stores them in {\tt f}.\\

\item \verb?void rate_pn(struct relicparam* paramrelic, double f[], double r[], double T9)?\\
\\
This procedure computes the decay rate and reverse rate of $p \leftrightarrow n$ reactions, and stores them in {\tt f} and {\tt r} respectively, at a temperature {\tt T9} (in unit of $10^{9}$ K).\\

\item \verb?void rate_all(double f[], double T9)?\\
\\
This procedure computes all the other reactions (summarized in Appendix \ref{nuclnet}) at temperature {\tt T9} and stores them in {\tt f}.\\

\item \verb?int nucl(struct relicparam* paramrelic, double *eta, double *H2_H, double *He3_H2,?
\verb?double *Yp, double *Li7_H, double *Li6_Li7)?\\
\\
This function is the main routine of the program, as it calculates using the \verb?paramrelic? structure the baryon-to-photon ratio \verb?eta?, the ratio of the element abundances \verb?H2_H?, \verb?He3_H2?, \verb?Li7_H?, \verb?Li6_Li7?, and the helium abundance \verb?Y_p?. It return 0 if the calculation failed, or 1 otherwise.\\

\item \verb?int bbn_excluded(struct relicparam* paramrelic)?\\
\\
This function is a ``container'' function which calls the routine \verb?nucl? and returns 0 if the abundances satisfy the constraints of Appendix~\ref{constraints}, 1 otherwise, and -1 if the calculation failed.

\end{itemize}
%
\section{Compilation and installation instructions}%
\label{compilation}
\verb?AlterBBN? is written for a C compiler respecting the C99 standard, and it has been tested successfully with the GNU C and the Intel C Compilers on Linux and Mac. The package can be downloaded from:\\
\\
{\tt http://superiso.in2p3.fr/relic/alterbbn}\\
\\
After unpacking, the following main directory is created:\\
\\
\verb?alterbbn_vX.X?\\
\\
This directory contains the \verb?src/? directory, in which all the source files can be found. The main directory contains also a \verb?Makefile?, a \verb?README?, five sample main programs (\verb?stand_cosmo.c?, \verb?alter_eta.c?, \verb?alter_neutrinos.c?, \verb?alter_standmod.c? and \verb?alter_reheating.c?). The compilation options should be defined in the \verb?Makefile?, and in particular the C compiler name and its specific flags, if needed.\\
Additional information can be found in the \verb?README? file.\\
To compile the library, type\\
\\
\verb?make?\\
\\
\noindent This creates \verb?libbbn.a? in \verb?src/?. Then, to compile one of the five programs provided in the main directory, type\\
\\
\verb?make name    ? or \verb?    make name.c?\\
\\
\noindent where \verb?name? can be \verb?stand_cosmo?, \verb?alter_eta?, \verb?alter_neutrinos?, \verb?alter_standmod? or \verb?alter_reheating?. This generates an executable program with the \verb?.x? extension:\\
\\
\verb?stand_cosmo.x? calculates the abundance of the elements in the standard cosmological model, assuming the baryon-to-photon ratio measured by WMAP.\\
\\
\verb?alter_eta.x? computes the abundance of the elements in the standard cosmological model, with the baryon-to-photon ratio given by the user.\\
\\
\verb?alter_neutrinos.x? calculates the abundance of the elements in the standard cosmological model, and give the user the possibility to alter the number of neutrinos.\\
\\
\verb?alter_standmod.x? and \verb?alter_reheating? computes the abundance of the elements in cosmological scenarios with modified expansion rates and entropy contents.
%
\section{Input and output description}%
\label{sample}
In the following, we describe the input and output of the main programs.

\subsection{Standard cosmology}
The program \verb?stand_cosmo.x? computes the abundance of the elements in the standard cosmological model assuming the baryon-to-photon ratio $\eta=6.19 \times 10^{-10}$ measured by WMAP \cite{wmap}. No parameter is needed by this program.
Running the program with:\\
\verb?./stand_cosmo.x?\\
returns\\
\verb?Yp               H2/H            He3/H2          Li7/H           Li6/Li7?\\
\verb?2.448e-01        2.544e-05       4.079e-01       4.477e-10       2.429e-05?\\
\verb?Compatible with BBN constraints?

\subsection{Standard cosmology with $\eta$ modification}
The program \verb?alter_eta.x? computes the abundance of the elements in the standard cosmological model, and needs the baryon-to-photon ratio $\eta$ as input. Running the program with:\\
\verb?./alter_eta.x 3e-10?\\
returns\\
\verb?Yp               H2/H            He3/H2          Li7/H           Li6/Li7?\\
\verb?2.374e-01        7.967e-05       2.020e-01       1.168e-10       2.692e-04?\\
\verb?Excluded by BBN constraints?

\subsection{Standard cosmology with modified neutrino number}
The program \verb?alter_neutrinos.x? computes the abundance of the elements in the standard cosmological model with a modified neutrino number. It needs as input the number of neutrino families $N_\nu$.\\
Running the program with:\\
\verb?./alter_neutrinos.x 4.5?\\
returns\\
\verb?Yp               H2/H            He3/H2          Li7/H           Li6/Li7?\\
\verb?2.629e-01        3.053e-05       3.614e-01       3.914e-10       3.641e-05?\\
\verb?Excluded by BBN constraints?

\subsection{Effective energy and entropy densities}
The program \verb?alter_standmod.x? computes the abundance of the elements while adding to the standard cosmological model an effective energy density such that
\begin{equation}
\rho_D =  \kappa_\rho \rho_{rad}(T_{BBN}) \bigl(T/T_{BBN}\bigr)^{n_\rho} \;,
\end{equation}
and/or an effective entropy density
\begin{equation}
 s_D =  \kappa_s s_{rad}(T_{BBN}) \bigl(T/T_{BBN}\bigr)^{n_s} \;,
\end{equation}
which modify the Early Universe properties without having observational consequences if chosen adequately \cite{arbey,arbey2}. A description of the model and of the related equations can be found in Sections \ref{modexprate} and \ref{modentropy}. The necessary arguments to this program are:
\begin{itemize}
 \item $\kappa_\rho$: ratio of dark energy density over radiation energy density at BBN time,\vspace*{-0.2cm}
 \item $n_\rho$: dark energy density decrease exponent,\vspace*{-0.2cm}
 \item $\kappa_s$: ratio of dark entropy density over radiation entropy density at BBN time,\vspace*{-0.2cm}
 \item $n_s$: dark entropy density decrease exponent.
\end{itemize}
Optional arguments can also be given:
\begin{itemize}
 \item $T_\rho$: temperature in GeV below which the dark energy density is set to 0,
 \item $T_s$: temperature in GeV below which the dark entropy density is set to 0.
\end{itemize}
For example, the command
\verb?./alter_standmod.x 0.1 6 1.e-3 5?\\
returns\\
\verb?Yp               H2/H            He3/H2          Li7/H           Li6/Li7?\\
\verb?2.315e-01        4.558e-05       2.831e-01       1.933e-10       9.173e-05?\\
\verb?Compatible with BBN constraints?

\subsection{Effective reheating model}
The program \verb?alter_standmod.x? computes the abundance of the elements while adding to the standard cosmological model an effective energy density such that
\begin{equation}
\rho_D =  \kappa_\rho \rho_{rad}(T_{BBN}) \bigl(T/T_{BBN}\bigr)^{n_\rho} \;,
\end{equation}
and/or an effective entropy production
\begin{equation}
\Sigma_D =  \kappa_\Sigma \Sigma_{rad}(T_{BBN}) \bigl(T/T_{BBN}\bigr)^{n_\Sigma} \;,
\end{equation}
which modify the Early Universe properties without having observational consequences if chosen adequately \cite{arbey3}. A description of the model and of the related equations can be found in Sections \ref{modexprate} and \ref{modentropy}. Note that $\Sigma_D$ is related to $s_D$ by Eq. (\ref{sigmas}). The necessary arguments to this program are:
\begin{itemize}
 \item $\kappa_\rho$: ratio of dark energy density over radiation energy density at BBN time,\vspace*{-0.2cm}
 \item $n_\rho$: dark energy density decrease exponent,\vspace*{-0.2cm}
 \item $\kappa_\Sigma$: ratio of dark entropy production over radiation entropy production at BBN time,\vspace*{-0.2cm}
 \item $n_\Sigma$: dark entropy production exponent,
 \item $T_r$: temperature in GeV below which the dark energy density and the entropy production are set to 0.
\end{itemize}
For example, the command
\verb?./alter_reheating.x 0.1 6 0.1 -1 1e-3?\\
returns\\
\verb?Yp               H2/H            He3/H2          Li7/H           Li6/Li7?\\
\verb?2.561e-01        2.615e-05       4.007e-01       4.590e-10       2.583e-05?\\
\verb?Compatible with BBN constraints?\\
\\
\noindent In order to perform scans in different cosmological models, the user is invited to write his/her own programs using the above main programs as guidelines.\\
\section{Results}
\label{result}

\verb?AlterBBN? has been thoroughly tested, and provides results in agreement with those of NUC123 \cite{kawano}.

\begin{figure}[ht!]
\centering
\includegraphics[width=8.05cm]{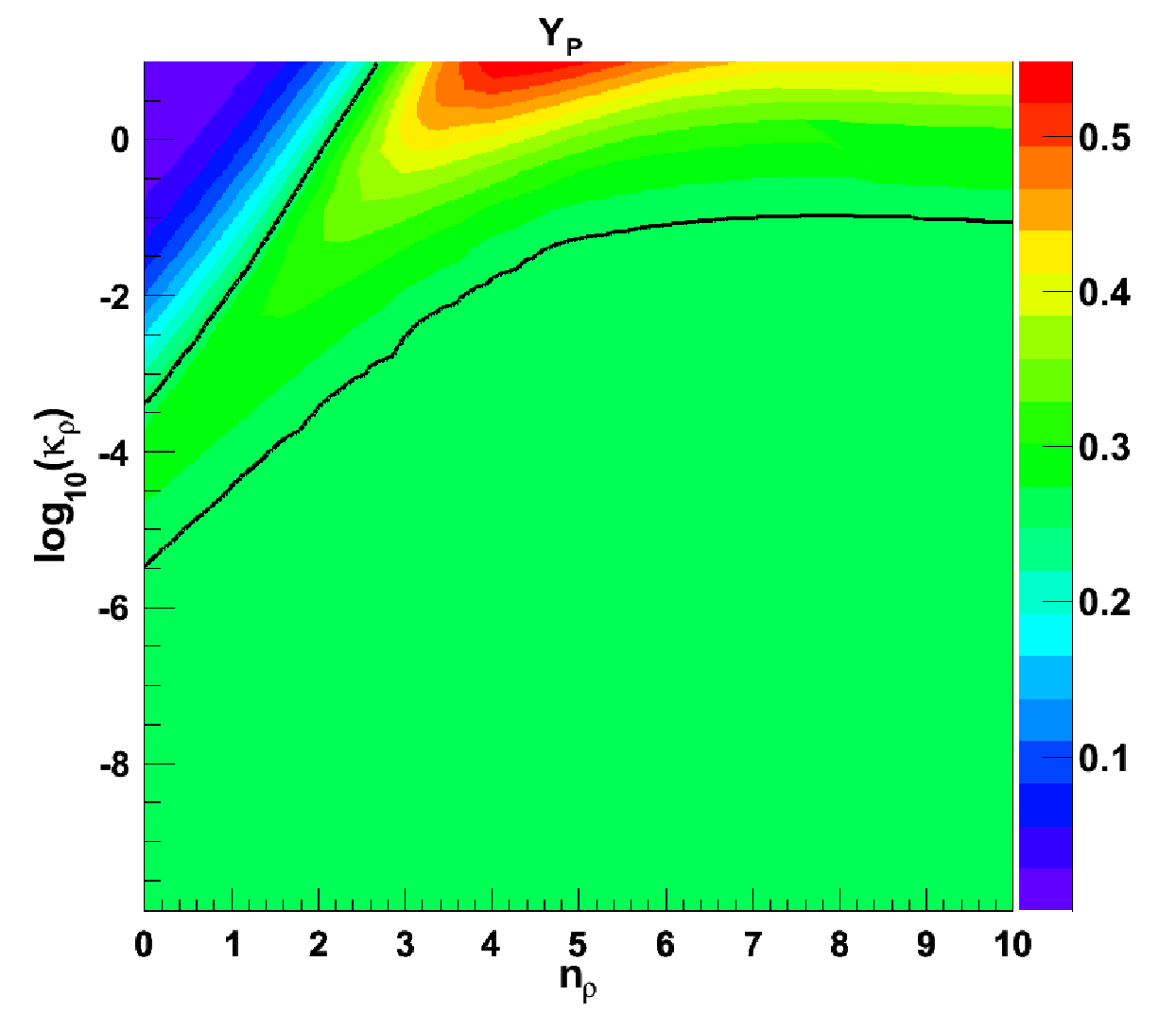}\includegraphics[width=8.05cm]{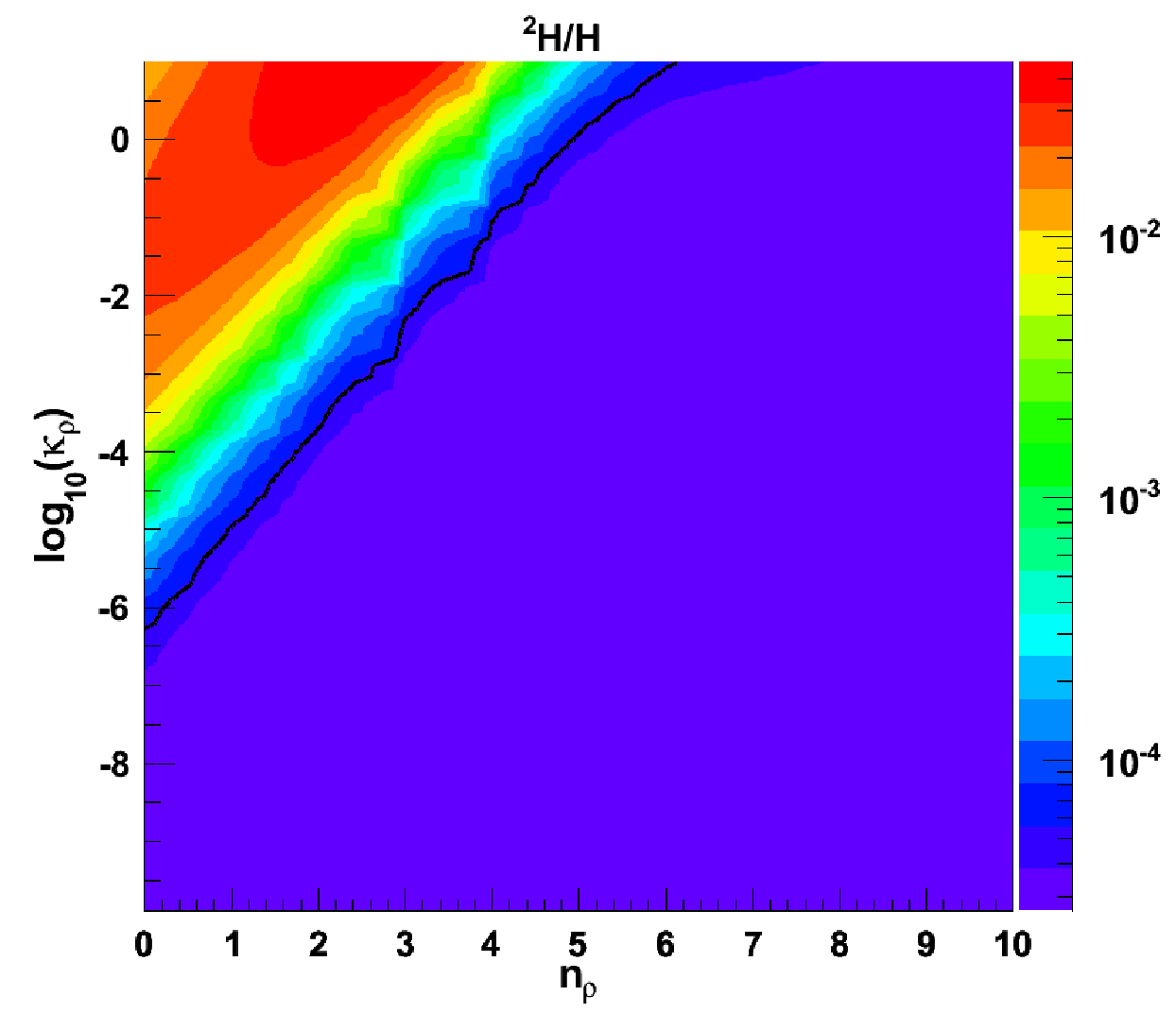}
\caption{Constraints from $Y_p$ (left) and $^2H/H$ (right) on the effective dark energy density parameters $(n_\rho,\kappa_\rho)$. The parameter regions excluded by BBN are located between the black lines for $Y_p$, and in the upper left corner for $^2H/H$. The colours correspond to different values of $Y_p$ and $^2H/H$.\label{BBN}}
\end{figure}

Considering the different cosmological models implemented, it is possible to perform scans on the cosmological parameters of these models to determine constraints. For example, in Fig.~\ref{BBN}, the current limits on the modified expansion properties from the $Y_p$ and $^2H/H$ BBN constraints are presented. The area between the back lines on the left plot and the area on the top of the black lines on the right plot lead to unfavored element abundances. The constraints of Appendix \ref{constraints} are used in this figure.

Also, \verb?AlterBBN? will soon be interfaced and provided in the SuperIso Relic package \cite{superisorelic}, so that the implemented cosmological models can be tested at the same time by BBN constraints as well as by particle physics constraints.

\newpage~

\newpage
\appendix
%
\append{Nuclear reaction network}
\label{nuclnet}
\verb?AlterBBN? currently includes a network of 88 nuclear reactions, which are gathered in the following table.\\
\\
\begin{table}[ht]
\hspace*{-1.cm}\mbox{\begin{tabular}{|c|c|c|}
\hline
 nb & ref & reaction\\
\hline\hline
0 & \cite{smith92} & $n \leftrightarrow p$\\
\hline\hline
1 & \cite{tilly87} & $^3\!H \rightarrow e^- + \nu + ^3\!He$\\
\hline
2 & \cite{ajzenberg88} & $^8\!Li \rightarrow e^- + \nu + 2\,^4\!He$\\
\hline
3 & \cite{ajzenberg90} & $^{12}\!B \rightarrow e^- + \nu + ^{12}\!C$\\
\hline
4 & \cite{ajzenberg86} & $^{14}\!C \rightarrow e^- + \nu + ^{14}\!N$\\
\hline
5 & \cite{ajzenberg88} & $^8\!B \rightarrow e^+ + \nu + 2\,^4\!He$\\
\hline
6 & \cite{ajzenberg90} & $^{11}\!C \rightarrow e^+ + \nu + ^{11}\!B$\\
\hline
7 & \cite{ajzenberg90} & $^{12}\!N \rightarrow e^+ + \nu + ^{12}\!C$\\
\hline
8 & \cite{ajzenberg86} & $^{13}\!N \rightarrow e^+ + \nu + ^{13}\!C$\\
\hline
9 & \cite{ajzenberg86} & $^{14}\!O \rightarrow e^+ + \nu + ^{14}\!N$\\
\hline
10 & \cite{ajzenberg86} & $^{15}\!O \rightarrow e^+ + \nu + ^{15}\!N$\\
\hline\hline
11 & \cite{smith92} & $H + n \rightarrow \gamma + ^2\!H$\\
\hline
12 & \cite{wagoner69} & $^2\!H + n \rightarrow \gamma + ^3\!H$\\
\hline
13 & \cite{wagoner69} & $^3\!He + n \rightarrow \gamma + ^4\!He$\\
\hline
14 & \cite{malaney93} & $^6\!Li + n \rightarrow \gamma + ^7\!Li$\\
\hline
15 & \cite{smith92} & $^3\!He + n \rightarrow p + ^3\!H$\\
\hline
16 & \cite{smith92} & $^7\!Be + n \rightarrow p + ^7\!Li$\\
\hline
17 & \cite{caughlan88} & $^6\!Li + n \rightarrow \alpha + ^3\!H$\\
\hline
18 & \cite{wagoner69} & $^7\!Be + n \rightarrow \alpha + ^4\!He$\\
\hline
19 & \cite{smith92} & $^2\!H + p \rightarrow \gamma + ^3\!He$\\
\hline
20 & \cite{smith92} & $^3\!H + p \rightarrow \gamma + ^4\!He$\\
\hline
21 & \cite{caughlan88} & $^6\!Li + p \rightarrow \gamma + ^7\!Be$\\
\hline
22 & \cite{caughlan88} & $^6\!Li + p \rightarrow \alpha + ^3\!He$\\
\hline
23 & \cite{smith92} & $^7\!Li + p \rightarrow \alpha + ^4\!He$\\
\hline
24 & \cite{caughlan88} & $^2\!H + \alpha \rightarrow p + ^6\!Li$\\
\hline
25 & \cite{smith92} & $^3\!H + \alpha \rightarrow p + ^7\!Li$\\
\hline
26 & \cite{smith92} & $^3\!He + \alpha \rightarrow p + ^7\!Be$\\
\hline
27 & \cite{smith92} & $^2\!H  + D \rightarrow p + ^3\!He$\\
\hline
28 & \cite{smith92} & $^2\!H  + D \rightarrow n + ^3\!H$\\
\hline
29 & \cite{smith92} & $^3\!H  + D \rightarrow n + ^4\!He$\\
\hline
 \end{tabular}
\hspace*{-0.cm}\begin{tabular}{|c|c|c|}
\hline
  nb & ref & reaction\\
\hline\hline
30 & \cite{caughlan88} & $^3\!He  + D \rightarrow p + ^4\!He$\\
\hline
31 & \cite{caughlan88} & $^3\!He + ^3\!He \rightarrow 2\,p + ^4\!He$\\
\hline
32 & \cite{caughlan88} & $^7\!Li  + D \rightarrow n + \alpha + ^4\!He$\\
\hline
33 & \cite{caughlan88} & $^7\!Be  + D \rightarrow p + \alpha + ^4\!He$\\
\hline
34 & \cite{wagoner69} & $^7\!Li + n \rightarrow \gamma + ^8\!Li$\\
\hline
35 & \cite{wagoner69} & $^{10}\!B + n \rightarrow \gamma + ^{11}\!B$\\
\hline
36 & \cite{malaney93} & $^{11}\!B + n \rightarrow \gamma + ^{12}\!B$\\
\hline
37 & \cite{caughlan88} & $^{11}\!C + n \rightarrow p + ^{11}\!B$\\
\hline
38 & \cite{caughlan88} & $^{10}\!B + n \rightarrow \alpha + ^7\!Li$\\
\hline
39 & \cite{caughlan88} & $^7\!Be + p \rightarrow \gamma + ^8\!B$\\
\hline
40 & \cite{caughlan88} & $^9\!Be + p \rightarrow \gamma + ^{10}\!B$\\
\hline
41 & \cite{caughlan88} & $^{10}\!B + p \rightarrow \gamma + ^{11}\!C$\\
\hline
42 & \cite{caughlan88} & $^{11}\!B + p \rightarrow \gamma + ^{12}\!C$\\
\hline
43 & \cite{caughlan88} & $^{11}\!C + p \rightarrow \gamma + ^{12}\!N$\\
\hline
44 & \cite{wagoner69} & $^{12}\!B + p \rightarrow n + ^{12}\!C$\\
\hline
45 & \cite{caughlan88} & $^9\!Be + p \rightarrow \alpha + ^6\!Li$\\
\hline
46 & \cite{caughlan88} & $^{10}\!B + p \rightarrow \alpha + ^7\!Be$\\
\hline
47 & \cite{wagoner69} & $^{12}\!B + p \rightarrow \alpha + ^9\!Be$\\
\hline
48 & \cite{caughlan88} & $^6\!Li + \alpha \rightarrow \gamma + ^{10}\!B$\\
\hline
49 & \cite{caughlan88} & $^7\!Li + \alpha \rightarrow \gamma + ^{11}\!B$\\
\hline
50 & \cite{caughlan88} & $^7\!Be + \alpha \rightarrow \gamma + ^{11}\!C$\\
\hline
51 & \cite{wagoner69} & $^8\!B + \alpha \rightarrow p + ^{11}\!C$\\
\hline
52 & \cite{malaney93} & $^8\!Li + \alpha \rightarrow n + ^{11}\!B$\\
\hline
53 & \cite{caughlan88} & $^9\!Be + \alpha \rightarrow n + ^{12}\!C$\\
\hline
54 & \cite{kawano} & $^9\!Be  + D \rightarrow n + ^{10}\!B$\\
\hline
55 & \cite{kawano} & $^{10}\!B  + D \rightarrow p + ^{11}\!B$\\
\hline
56 & \cite{kawano} & $^{11}\!B  + D \rightarrow n + ^{12}\!C$\\
\hline
57 & \cite{caughlan88} & $^4\!He + \alpha + n \rightarrow \gamma + ^9\!Be$\\
\hline
58 & \cite{caughlan88} & $^4\!He + 2\,\alpha \rightarrow \gamma + ^{12}\!C$\\
\hline
59 & \cite{kawano} & $^8\!Li + p \rightarrow n + \alpha + ^4\!He$\\[0.16cm]
\hline
 \end{tabular}
\hspace*{-0.cm}\begin{tabular}{|c|c|c|}
\hline
  nb & ref & reaction\\
\hline\hline
60 & \cite{kawano} & $^8\!B + n \rightarrow p + \alpha + ^4\!He$\\
\hline
61 & \cite{caughlan88} & $^9\!Be + p \rightarrow d + \alpha + ^4\!He$\\
\hline
62 & \cite{caughlan88} & $^{11}\!B + p \rightarrow 2\,\alpha + Be4$\\
\hline
63 & \cite{wagoner69} & $^{11}\!C + n \rightarrow 2\,\alpha + ^4\!He$\\
\hline
64 & \cite{wagoner69} & $^{12}\!C + n \rightarrow \gamma + ^{13}\!C$\\
\hline
65 & \cite{wagoner69} & $^{13}\!C + n \rightarrow \gamma + ^{14}\!C$\\
\hline
66 & \cite{wagoner69} & $^{14}\!N + n \rightarrow \gamma + ^{15}\!N$\\
\hline
67 & \cite{caughlan88} & $^{13}\!N + n \rightarrow p + ^{13}\!C$\\
\hline
68 & \cite{caughlan88} & $^{14}\!N + n \rightarrow p + ^{14}\!C$\\
\hline
69 & \cite{caughlan88} & $^{15}\!O + n \rightarrow p + ^{15}\!N$\\
\hline
70 & \cite{caughlan88} & $^{15}\!O + n \rightarrow \alpha + ^{12}\!C$\\
\hline
71 & \cite{caughlan88} & $^{12}\!C + p \rightarrow \gamma + ^{13}\!N$\\
\hline
72 & \cite{caughlan88} & $^{13}\!C + p \rightarrow \gamma + ^{14}\!N$\\
\hline
73 & \cite{caughlan88} & $^{14}\!C + p \rightarrow \gamma + ^{15}\!N$\\
\hline
74 & \cite{caughlan88} & $^{13}\!N + p \rightarrow \gamma + O14$\\
\hline
75 & \cite{caughlan88} & $^{14}\!N + p \rightarrow \gamma + ^{15}\!O$\\
\hline
76 & \cite{caughlan88} & $^{15}\!N + p \rightarrow \gamma + ^{16}\!O$\\
\hline
77 & \cite{caughlan88} & $^{15}\!N + p \rightarrow \alpha + ^{12}\!C$\\
\hline
78 & \cite{caughlan88} & $^{12}\!C + \alpha \rightarrow \gamma + ^{16}\!O$\\
\hline
79 & \cite{wagoner69} & $^{10}\!B + \alpha \rightarrow p + ^{13}\!C$\\
\hline
80 & \cite{caughlan88} & $^{11}\!B + \alpha \rightarrow p + ^{14}\!C$\\
\hline
81 & \cite{caughlan88} & $^{11}\!C + \alpha \rightarrow p + ^{14}\!N$\\
\hline
82 & \cite{caughlan88} & $^{12}\!N + \alpha \rightarrow p + ^{15}\!O$\\
\hline
83 & \cite{caughlan88} & $^{13}\!N + \alpha \rightarrow p + ^{16}\!O$\\
\hline
84 & \cite{caughlan88} & $^{10}\!B + \alpha \rightarrow n + ^{13}\!N$\\
\hline
85 & \cite{caughlan88} & $^{11}\!B + \alpha \rightarrow n + ^{14}\!N$\\
\hline
86 & \cite{wagoner69} & $^{12}\!B + \alpha \rightarrow n + ^{15}\!N$\\
\hline
87 & \cite{caughlan88} & $^{13}\!C + \alpha \rightarrow n + ^{16}\!O$\\
\hline
&&\\[0.65cm]
\hline
\end{tabular}}
\caption{Network of nuclear reactions implemented in {\tt AlterBBN}.}
\end{table}
%
\append{BBN constraints}%
\label{constraints}
The following conservative constraints \cite{jedamzik06} are used in the function \verb?bbn_excluded?:
\begin{eqnarray}
0.240 < Y_p < 0.258\;, \qquad 1.2 \times 10^{-5} < \!~^2\!H/H < 5.3 \times 10^{-5}\;,\qquad\qquad\label{BBNconstraints}\\
\nonumber 0.57 < \!~^3\!H/\!~^2\!H < 1.52\;, \qquad  \!~^7\!Li/H > 0.85 \times 10^{-10}\;,\qquad \!~^6\!Li/\!~^7\!Li < 0.66\;,
\end{eqnarray}
for the helium abundance $Y_p$ and the primordial $^2\!H/H$, $^3\!H/\!~^2\!H$, $^7\!Li/H$ and $^6\!Li/\!~^7\!Li$ ratios.\\
It is possible for the user to change these constraints in routine \verb?bbn_excluded? which can be found in \verb?src/bbn.c?.


\end{document}